\documentclass[runningheads]{llncs}
\usepackage{graphicx}
\usepackage{array}
\usepackage{subfigure}
\usepackage{setspace}
\usepackage{amssymb}
\usepackage{ulem}
\usepackage{booktabs}
\usepackage{amsmath}
\usepackage{xcolor}
\usepackage{multirow}
\usepackage{amsmath,amssymb,amsfonts}
\usepackage{textcomp}
\usepackage{algorithm}
\usepackage{algpseudocode} 
\usepackage[pagebackref=true,breaklinks=true,colorlinks=true,bookmarks=false,citecolor=blue,urlcolor=blue,linkcolor=blue]{hyperref}
\usepackage[misc]{ifsym}

\begin{document}
%
\title{Test-time Adaptation with Calibration of \\ Medical Image Classification Nets for \\ Label Distribution Shift}
%
\titlerunning{Test-time Adaptation for Label Distribution Shift}
%
\author{Wenao Ma\textsuperscript{1}, Cheng Chen\textsuperscript{1}, Shuang Zheng\textsuperscript{2,3}, Jing Qin\textsuperscript{4}, \\ Huimao Zhang\textsuperscript{2,3(\Letter)}, Qi Dou\textsuperscript{1(\Letter)}}
\institute{$^{1}$ Dept. of Computer Science and Engineering, The Chinese University of Hong Kong \\
$^{2}$ Department of Radiology, The First Hospital of Jilin University \\
$^{3}$ Jilin Provincial Key Laboratory of Medical Imaging \& Big Data \\
$^{4}$ Centre for Smart Health, The Hong Kong Polytechnic University}
\authorrunning{Ma et al.} 
%


%
\maketitle              
%
\begin{abstract}
Class distribution plays an important role in learning deep classifiers. When the proportion of each class in the test set differs from the training set, the performance of classification nets usually degrades.
Such a label distribution shift problem is common in medical diagnosis since the prevalence of disease vary over location and time. 
In this paper, we propose the first method to tackle label shift for medical image classification, which effectively adapt the model learned from a single training label distribution to arbitrary unknown test label distribution. 
Our approach innovates distribution calibration to learn multiple representative classifiers, which are capable of handling different one-dominating-class distributions. When given a test image, the diverse classifiers are dynamically aggregated via the consistency-driven test-time adaptation, to deal with the unknown test label distribution. 
We validate our method on two important medical image classification tasks including liver fibrosis staging and COVID-19 severity prediction. 
Our experiments clearly show the decreased model performance under label shift.
With our method, model performance significantly improves on all the test datasets with different label shifts for both medical image diagnosis tasks. Code is available at \href{https://github.com/med-air/TTADC}{https://github.com/med-air/TTADC}.

\end{abstract}
\section{Introduction}
Intelligent medical image diagnosis has witnessed great success 
on accurate predictions for various tasks such as disease staging~\cite{liu2016relationship,ren2018adversarial}, lesion diagnosis~\cite{hussein2019lung,mesejo2016computer}, and severity prediction~\cite{konwer2021attention,roy2020deep}.
However, real-world use of classification models is challenged by the inevitable shift in class distributions on test data at deployment~\cite{azizzadenesheli2019regularized,subbaswamy2020development,wu2021online,zhang2013domain}.
Usually, the proportion of samples belonging to each class is associated with
patient demographics and region-related prevalence of disease, which differs from one hospital to another.
This issue is called \textit{label distribution shift}, which means that the label distribution
can change across training and test datasets.
As label distribution plays a vital role in classification tasks~\cite{galar2011review,moreno2012unifying}, such shift can make the learned classifier become suboptimal on unseen datasets, thus suffering from performance degradation in testing.





Label distribution shifts are very common in medical diagnosis as the disease distributions vary across location and time. 
For example, the prevalence of liver diseases significantly differs among regions due to the difference in vaccination coverage \cite{williams2006global}. 
Such label shifts often degrade the performance of a learned classifier on test data, leading to erroneous predictions as observed in prior works~\cite{challen2019artificial,chen2021probabilistic,davis2017calibration}.   
For example, Davis et al. find that the prediction accuracy of their machine learning models decreases due to the declining incidence of acute kidney injury over time~\cite{davis2017calibration}.
Since the proportion of normal and disease cases differs between the screening and diagnostic scenarios, an accurate model for screening purpose could perform poorly for diagnosis purpose, even for the same disease~\cite{challen2019artificial}.  
Park et al.~\cite{park2021reliable} show in three disease classification models that dataset shifts including the label shift can lead to unreasonable predictions. 
Despite being observed in many real applications, the problem of label distribution shift has not yet been tackled for medical image diagnosis, severely hindering the large-scale deployment of deep models in clinical practice.

To generalize model under label shift, if the label distribution of test data can be known, such as the uniform distribution assumption made in \cite{ren2020balanced,wang2020long}, the label shift can be alleviated by re-sampling training data or adjusting the prediction probability in the softmax loss~\cite{peng2020large,ren2020balanced} accordingly.
In practical scenarios, however, it is unlikely to anticipate the label distribution of test data, which is usually unknown and arbitrary, and may even continuously change. 
In this regard, we aim to mitigate label shift in a highly practical yet challenging setting, where the test label distribution is unknown and the trained model itself must accommodate label shift by utilizing the test data only. 
To tackle this problem, we consider two key ingredients. 
Firstly, since the test label distribution can be arbitrary, it is important to enlarge the capacity of models for an extensive label distribution space. 
The difficulty lies in how to establish such a representative space during model learning from the training set with a fixed label distribution. 
Secondly, motivated by the recent test-time learning works~\cite{sun2020test,wang2020tent}, although the knowledge of test dataset is unknown during model training, it can be explored from the test data at inference time. 

In this paper, to our best knowledge, we present the first work to effectively tackle the label distribution shift in medical image classification. 
Our method learns representative classifiers with distribution calibration, by extending the concept of balanced softmax loss~\cite{ren2020balanced,zhang2021test} to simulate multiple distributions that one class dominates other classes. 
Compared with \cite{zhang2021test}, our method can be more flexible and be more targeted for ordinal classification, as our one-dominating-class distributions can represent more diverse label distributions and we use ordinal encoding instead of one-hot encoding to train the model.
Then, at model deployment to new test data, we dynamically combine the representative classifiers by adapting their outputs to the label distribution of test data.
The test-time adaptation is driven by a consistency regularization loss to adjust the weights of different classifier.
We evaluate our method on two important medical applications of liver fibrosis staging and COVID-19 severity prediction. With our proposed method, the label shift can be largely mitigated with consistent performance improvement. 

\section{Method}
\begin{figure}[!t]
    \centering
	\includegraphics[width =12.0cm]{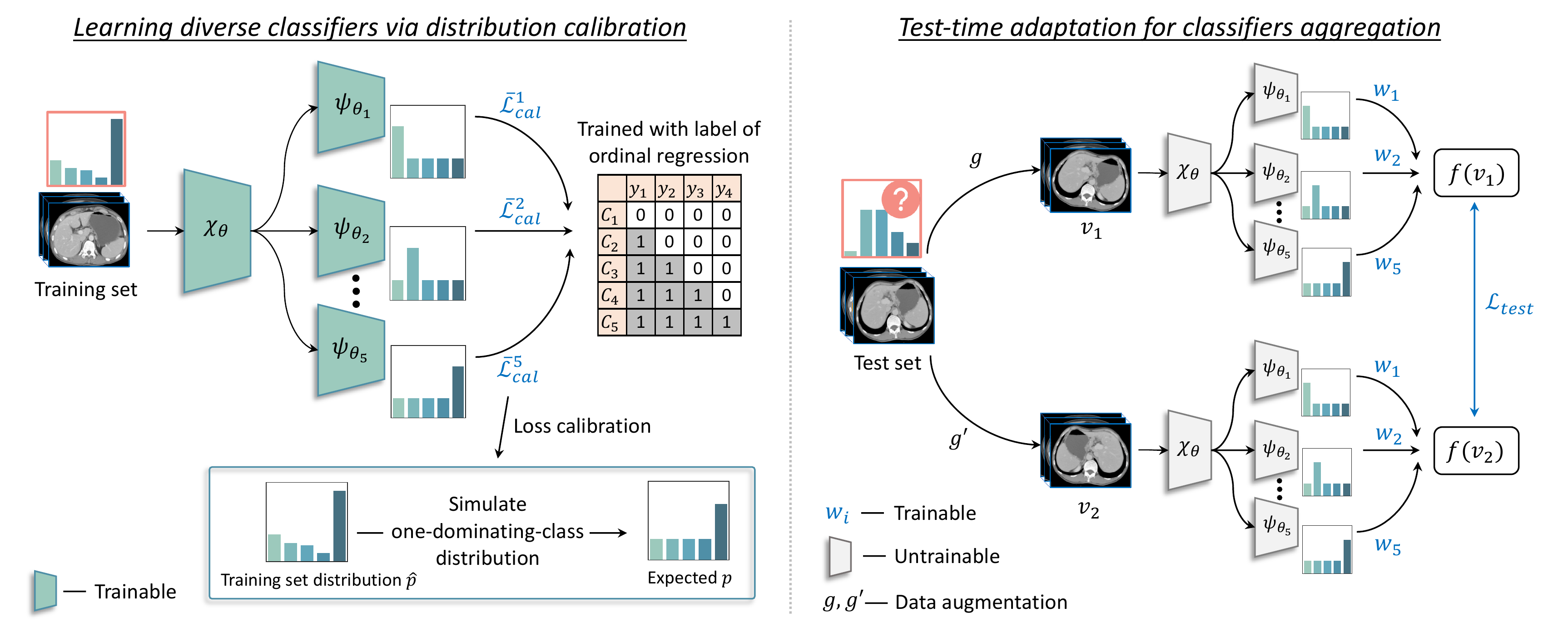}
	\vspace{-3mm}
	\caption{Overview of our proposed method for test-time adaptation by calibration of medical image classification networks for label distribution shift.}
	\label{pipeline}
\end{figure}
\subsection{Problem Formulation of Label Distribution Shift}
For disease diagnosis, consider a classification task that aims to train a model to predict the disease class $y$ correctly given an input image $x$. 
Let $\hat p(x,y)$ and $\tilde p(x,y)$ denote the training and test set distributions respectively.
In practice, a deployed model often suffers from label distribution shift, which means the label distribution of training set $\hat p(y)$ is different from that of test set $\tilde p(y)$, i.e., $\hat p(y)\neq \tilde p(y)$, but the conditional distributions are consistent, i.e., $\hat p(x|y)=\tilde p(x|y)$. 
This phenomenon is especially common in medical image classification, where the disease label $y$ is often the causal variable and the image data $x$ can be regarded as the manifestations of a disease \cite{scholkopf2012causal,wu2021online}. 
According to the Bayesian inference $\hat p(y|x) \! = \! \frac{\hat p(x|y)\hat p(y)}{ \hat p(x)}$, the model prediction $\hat p(y|x)$ is strongly coupled with the label distribution $\hat p(y)$, thus the shift in $\hat p(y)$ can cause erroneous prediction of $\hat p(y|x)$.
Regarding this problem, our goal is to adapt a classifier that is learned from the training set to perform well on any unseen test set with label distribution shift. 





\vspace{-0.3cm}
\subsection{Learning Diverse Classifiers via Distribution Calibration}
Since the test label distribution can be arbitrary, to generalize models under label shift, we consider it is important to enlarge the capacity of classifiers to a broad range of label distributions.
However, during training, the model is only presented to a fixed training label distribution thus has limited capacity. 
Inspired by balanced softmax~\cite{ren2020balanced} which calibrates skewed label distribution to be uniform by adding a compensating term to the softmax loss, we propose to learn diverse classifiers via dedicated distribution calibration. 
As shown in Fig.~\ref{pipeline}, our insight is to simulate representative one-dominating-class distributions so that the proper combination of learned classifiers can handle arbitrary test label distribution. 

Before introducing how to achieve distribution calibration, we first clarify the ordinal encoding in our classification task. 
To encourage classification network to learn the commonness of all classes and the distinctions between different classes, we use ordinal encoding~\cite{niu2016ordinal} instead of one-hot encoding for the ordinal classes in our liver fibrosis staging and COVID-19
severity prediction tasks.
This ordinal encoding performs multiple binary classifications with sigmoid function and combines the multiple binary outputs by taking the highest class that is predicted as 1 as the final prediction.
Furthermore, for distribution calibration in our ordinal regression, we extend the balanced softmax to the sigmoid function and derive the corresponding compensating term. 
Let $p(y_i=1|x)$ be the desired conditional probability for the expected label distribution, and $\hat{p}({y}_{i}=1|x)$ be the desired conditional probability of the training set, and assume $p({y}_{i}=1|x)$ is expressed by the standard sigmoid function of the network output $\phi_{i}$ in $i$-th ordinal vector: $p({y}_{i}=1|x) = \frac{e^{\phi_{i}}}{1+e^{\phi_{i}}}$, then the $\hat{p}({y}_{i}=1|x)$ with the same output $\phi_{i}$ can be expressed as:
\vspace{-3mm}
\begin{equation}
\label{equ2}
\hat{p}({y}_{i}=1|x)=\frac{e^{\phi_{i}-\log \left(\frac{{r'_{i}} }{1-{r'_{i}}}\cdot\frac{1-r_{i} }{r_{i}} \right)}}{1+e^{\phi_{i}-\log \left(\frac{{r'_{i}} }{1-{r'_{i}}}\cdot\frac{1-r_{i} }{r_{i}} \right)}},
\vspace{-0.1cm}
\end{equation}
where $r'_{i} $ and $r_{i} $ are the positive label proportion in the $i$-th ordinal vector for the expected label distribution $p$ and factual label distribution respectively $\hat{p}$, and the term $\log\left(\frac{{r'_{i}} }{1-{r'_{i}}}\cdot\frac{1-r_{i} }{r_{i}} \right)$ is the compensating term. The proof of Eq.(\ref{equ2}) is provided in the supplementary material. 

In this way, the calibrated loss function is: 
\vspace{-1mm}
\begin{equation}
\label{equ3}
\bar{\mathcal{L}}_{cal} = -\sum_{i=1}^{K-1} \left( y_{i}\log_{}{\hat{p}({y}_{i}=1|x)} +(1-y_{i})\log_{}{(1-\hat{p}({y}_{i}=1|x)} )\right),
\vspace{-1mm}
\end{equation}
Where $K$ denotes the total number of classes. This calibrated loss function enables the model learned on the training label distribution to generate the prediction for the expected label distribution.

Moreover, we aim to properly construct different $r'_{i}$ to simulate $K$ one-dominating-class distributions for $K$ classifiers. Assume the proportion of dominating class $j$ is $\lambda$ times other classes, then the value of
${r'_{i}}$ can be calculated as:
\vspace{-0mm}
\begin{equation}
\label{equ4}
{r'_{i}} = 1-\frac{i - \textbf{1}_{i \ge j}\cdot(1-\lambda)}{\lambda + K - 1},
\vspace{-0cm}
\end{equation}
where $\textbf{1}_{i \ge j}$ is the indicator function. The derivation of Eq.(\ref{equ4}) can be found in the supplementary material. 
Notably, our distribution-calibrated networks use independent parameters only at the last stages and fully-connected layer of networks, while share the parameters at other layers (see the shared network $\chi _{\theta}$ and the independent networks $\psi_{\theta}$ in Fig.~\ref{pipeline}).
This is motivated by the observation that decoupling the representation learning and classification gives more generalizable representations~\cite{kang2019decoupling}.
In this way, we obtain diverse classifiers to handle different label distributions, but adding only minimal computational cost.

\vspace{-0.3cm}
\subsection{Test-time Adaptation for Dynamic Classifier Aggregation}
After obtaining diverse distribution-calibrated classifiers during training phase, then at test time, the key is how to aggregate these classifiers to handle the unknown test label distribution with the given inference samples. 
To build the connection between the obtained classifiers and the test data, we aggregate the outputs of all classifiers with learnable weights, which are dynamically adapted using information implicitly provided by the test data. It's worth to mention that a set of test data, which can reflect the label distribution in the test center, should be accessible simultaneously during this phase.  

Specifically, the aggregated output is defined as $\hat{p}_{\text{agg}}=\sum_{k=1}^{K} w_{k}\hat{p}_{k}$, where $\sum_{k=1}^{K} w_{k}=1$ and $\hat{p}_{k}$ is the output of $k$-th classifiers with the form of ordinal vector.
As different combination of $\{w_{1}, w_{2}, ..., w_{K}\}$ can enable the model to deal with different test label distributions, the aim of our test-time adaptation is to find the optimal combination for a given test set. 
Our assumption is that if the aggregated model has adapted to a particular test label distribution, for the test images generated from such a label distribution, the model should give similar predictions to perturbed versions of the same image.
Based on this assumption, we design a consistency regularization mechanism to drive the test-time learning. 
Given an input $x$, we generate two augmented views $g(x)=v_{1}$ and $g'(x)=v_{2}$ using the data augmentation approaches, including rotating, flipping, and shifting the images, and adding Gaussian noise to the images. The two views are then forwarded to the trained model $f(\cdot)$ respectively, yielding the ordinal encoded output $f(v_{1})=\hat{p}_{\text{agg}}=w_{1}\cdot\hat{p}_{1}+w_{2}\cdot\hat{p}_{2}+\dots +w_{K}\cdot\hat{p}_{K}$ and $f(v_{2})=\hat{p}'_{\text{agg}}=w_{1}\cdot\hat{p}'_{1}+w_{2}\cdot\hat{p}'_{2}+\dots +w_{K}\cdot\hat{p}'_{K}$. 
The consistency regularization for the outputs of the two views is imposed with a cosine similarity loss:
\vspace{-0.1cm}
\begin{equation}
\label{equ5}
\mathcal{L}_{\text{test}}=-cos(f(v_{1}),f(v_{2}))=-\frac{f(v_{1}) \cdot f(v_{2})}{\left \| f(v_{1}) \right \|_{2}\times  \left \| f(v_{2}) \right \|_{2} },
\vspace{-0.1cm}
\end{equation}
The loss $\mathcal{L}_{\text{test}}$ drives the updates of the weights set $\{w_{1}, w_{2}, ..., w_{K}\}$ with the implicit knowledge of label distribution on the test set, while the other network parameters of $f(\cdot)$ are frozen. This implicit knowledge is reflected by the consistency that measures whether the aggregated model has adapted to the test label distribution successfully.
Each weight of $\{w_{1}, w_{2}, ..., w_{K}\}$ is initialized to $\frac{1}{K}$ and we use softmax function to maintain the sum of them equals to one after each iteration. As a result, the test results can be obtained after the test-time adaptation given the optimized weights set.

\section{Experiment}
\subsection{Dataset and Experimental Setup}
\textbf{Datasets.} We have validated our proposed method on two tasks: 1) liver fibrosis staging with an in-house abdominal CT dataset, and 2) COVID-19 severity prediction with a public chest CT dataset (iCTCF~\cite{ning2020open}). 
The liver CT dataset consists of three centers with different label distributions, including 823 cases from our center, 99 cases from external center A and 50 cases from external center B. The ground truths of the liver fibrosis staging come from the pathology results of liver biopsy. The liver fibrosis disease is divided into 5 stages, including no fibrosis (F0), portal fibrosis without septa (F1), portal fibrosis with few septa (F2), numerous septa without cirrhosis (F3) and cirrhosis (F4). 
Segmentation of the liver is pre-computed with an out-of-the-box tool in a related clinical study, so we adopt it in our paper as the region of interest for classification.
The slice thickness of the CT images is 5 $mm$ and the in-plane resolution is 512 $\times$ 512. 
For the COVID-19 dataset, it contains 969 cases from HUST-Union Hospital for training and 370 cases from HUST-Liyuan Hospital for test. The severity of COVID-19 is divided to 6 levels: control (S0), suspected (S1), mild (S2), regular (S3), severe (S4) and critically (S5). The preprocessing and automatic lung segmentation process are the same as a recent work~\cite{bao2022covid} on this dataset. 


\textbf{Experimental setting.} For liver fibrosis staging, we take 630 cases from our center as the training set, 193 cases from our center as evaluation set and the data from two external centers as two different test sets. For COVID-19 severity prediction, we use the data from HUST-Union Hospital for training and data from HUST-Liyuan Hospital for test.
Label distribution statistics of different centers for both datasets are provided in supplementary.

\textbf{Evaluation metrics.} For both tasks, the diagnosis performance is evaluated with accuracy, area under the receiver operating characteristic curve (AUC) and Obuchowski index (OI) \cite{obuchowski2001assessing}, as reported in related works \cite{bao2022covid,choi2018development,park2019radiomics}. 
Considering the AUC is defined for binary classification while ours are multi-class classification tasks, we combine the classes and convert the multi-class classification to several binary classifications. Specifically, we calculate the AUC of F0 vs F1-4, F0-1 vs F2-4, F0-2 vs F3-4 and F0-3 vs F4 for the liver fibrosis staging, and the AUC of S0 vs S1-5, S0-1 vs S2-5, S0-2 vs S3-5, S0-3 vs S4-5 and S0-4 vs S5 for COVID-19 severity prediction.
We report the average of all the AUC values as overall performance. The Obuchowski index (OI) is a metric which is proved to have no bias when label distributions are different between training and test sets \cite{lambert2008measure}. 

\begin{figure}[!t]
    \centering
	\includegraphics[width =0.95\textwidth]{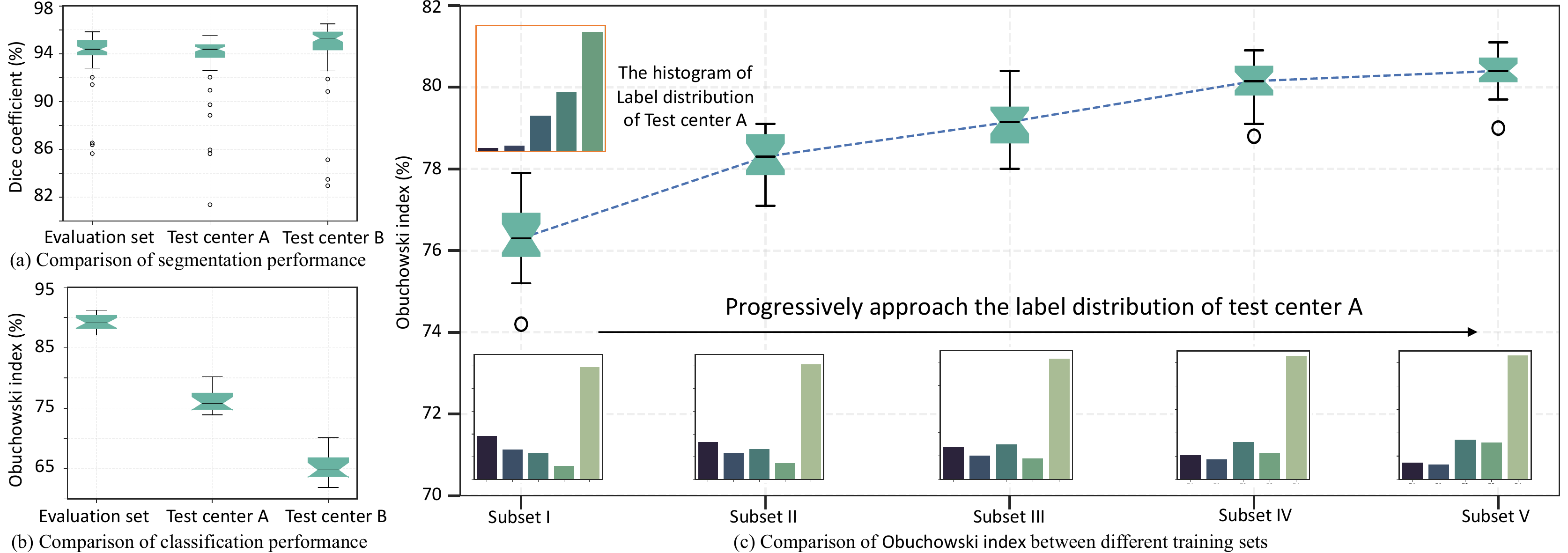}
	\caption{Analysis of model performance with label distribution shift.} 
	\label{exp}
\end{figure}
\textbf{Implementation details.} 
Considering model efficiency while still capturing 3D information in CT scans, we use ResNet-50 to get a vector of spatial features and then forward the features of adjacent slices to a LSTM module and a fully connected layer for classification.
We train the models using Adam with an initial learning rate of $1e-5$, a weight decay of $1e-4$ and batch size of 4. Our models are implemented using a workstation with four NVIDIA TITAN Xp GPUs.


\subsection{Experimental Results} 
\subsubsection{Observation of label distribution shift.} 
Label distribution shift and data distribution shift are two types of dataset shift, as introduced in previous work \cite{subbaswamy2020development}. We first clearly show in the multi-center liver CT datasets that under label distribution shift, the performance of classification model would degrade. 
Fig.~\ref{exp} (a) and (b) present that the segmentation performance for region-of-interest liver extraction is consistent between the evaluation set and test sets, while the final classification performance of Obuchowski index largely decreases by 12.9\% and 27.5\% at test set A and B. It worth to mention that the label distribution of evaluation set is consistent with the training set while the test sets are not.
In Fig.~\ref{exp} (c), we progressively adjust the class distribution of training set to approach the label distribution of test set A, by random sampling a certain proportion of images belonging to each class. We can see that the classification performance increases when the class distribution of the training set becomes closer to the test set. 
These experiments clearly demonstrate it is indeed the label shift causes the performance drop of classification model in our datasets.

\subsubsection{Comparison with state-of-the-art methods.} 
We here compare our method with state-of-the-art approaches for label shift in natural images as strong competitors, including \textbf{BALMS} \cite{ren2020balanced}, which calibrates the training label distribution to be uniform, \textbf{LADE} \cite{hong2021disentangling}, which disentangles the training label distribution from the model prediction, and \textbf{TADE} \cite{zhang2021test}, which also proposes to train multiple networks with different expertise but their networks are less representative than ours. 
Note that BALMS, LADE, and TADE need to use our derived compensating term in Eq.~\ref{equ2} to be applied in our classification tasks with ordinal regression. 
We also compare our method with \textbf{TENT} \cite{wang2020tent}, which is a general test-time adaptation approach for domain shift problem, and \textbf{Focal Loss} \cite{lin2017focal}, which can alleviate class imbalance by increasing the focus on hard samples. 

Table~\ref{tab4} presents the comparison results on the test centers of both liver fibrosis staging and COVID-19 severity prediction.
Our TTADC significantly improves the model performance over baseline on all test sets, with 4.6\%, 3.6\%, 2.7\% increase in AUC, 8.5\%, 10.0\%, 4.0\% increase in Accuracy, and 3.9\%, 3.3\%, 4.6\% increase in OI respectively, outperforming all the comparison methods. 
The results validate the effectiveness of our distribution calibration and test-time adaptation on addressing arbitrary label shift. 
Our method clearly outperforms the domain adaptation method TENT, showing the necessity of designing approach specifically for label shift.
Although not significant, Focal loss can also generally improve over baseline, indicating the alleviation of class imbalance may help reduce the effect of label shift.
The other methods on tackling label shift generally outperform TENT and Focal loss. 
Our method and TADE which learn multiple classifiers obtain better performance than BALMS and LADE which use uniform distribution assumption, showing the importance of enlarging the model capacity for proper test-time adaptation. 
Our method also clearly outperforms TADE, demonstrating the combination of our one-dominating-class distributions can represent more diverse test label distributions. 
\renewcommand{\arraystretch}{1.1}
\begin{table*}[t]
\centering
\caption{Quantitative comparison of different methods on the test sets of the two tasks. Results are reported with average and standard deviation over three independent runs. }
\resizebox{1.0\textwidth}{!}{%
	    \setlength\tabcolsep{2.0pt}
	    \scalebox{1.1}{
	    \begin{tabular}{l||ccc|ccc||ccc}
	    \hline
        \hline
        \multirow{3}{*}{Methods} 
        &\multicolumn{6}{c||}{\textbf{Task 1}: Liver fibrosis staging}  &\multicolumn{3}{c}{\textbf{Task 2}: COVID-19}  \\
        
        \cline{2-7}
        &\multicolumn{3}{c|}{Test center A} & \multicolumn{3}{c||}{Test center B} 
        &\multicolumn{3}{c}{severity prediction}\\
        \cline{2-10}
        
        &~~AUC~~ & Accuracy & ~~OI~~ &~~AUC~~ & Accuracy & ~~OI~~&~~AUC~~ & Accuracy & ~~OI~~  \\ 
        \hline 
        \hline

        Baseline & 77.7$\pm$0.7 & 52.5$\pm$0.8  & 76.3$\pm$0.5 & 68.8$\pm$0.6 & 40.7$\pm$0.9& 66.3$\pm$0.5 &68.4$\pm$0.8& 36.2$\pm$1.0 &65.2$\pm$0.6   \\ 
        \hline
        
        BALMS \cite{ren2020balanced} & 80.3$\pm$0.5 & 54.9$\pm$0.5 & 78.3$\pm$0.5  & 70.1$\pm$0.7  &44.0$\pm$1.6 & 67.0$\pm$0.4& 69.5$\pm$0.6& 36.2$\pm$1.0 & 66.4$\pm$0.5 \\
        LADE \cite{hong2021disentangling} & 80.6$\pm$0.5 & 57.6$\pm$0.8 & 78.5$\pm$0.5 &  69.3$\pm$0.6 &  46.0$\pm$1.6   & 67.9$\pm$0.5  & 68.3$\pm$0.6 & 37.2$\pm$0.9 & 66.2$\pm$0.5  \\
        TADE \cite{zhang2021test} & 80.9$\pm$0.6 & 59.9$\pm$1.9 & 79.2$\pm$0.5 & 70.2$\pm$0.8 & 47.3$\pm$0.9  & 68.5$\pm$0.7  & 69.6$\pm$0.8 & 38.3$\pm$1.6 & 68.4$\pm$0.6  \\
        
        TENT \cite{wang2020tent} & 78.9$\pm$0.8  & 53.2$\pm$0.5  &77.0$\pm$0.7  & 69.9$\pm$0.5  & 42.7$\pm$0.9  &67.1$\pm$0.5 &69.1$\pm$0.8 & 36.4$\pm$0.8 & 65.6$\pm$0.7   \\
        Focal Loss \cite{lin2017focal} & 80.2$\pm$0.6 & 53.2$\pm$0.5 & 78.0$\pm$0.5  & 69.1$\pm$0.7   & 43.3$\pm$0.9 & 67.7$\pm$0.6& 69.5$\pm$0.6 & 36.5$\pm$1.0 &66.5$\pm$0.5  \\ 
        
        \hline
        \textbf{TTADC (ours)} & \textbf{82.3$\pm$0.4}  &  \textbf{61.0$\pm$1.0 }& \textbf{80.2$\pm$0.4} & \textbf{72.4$\pm$0.6}  & \textbf{50.7$\pm$0.9} & \textbf{69.6$\pm$0.4} &\textbf{71.1$\pm$0.6} &\textbf{40.2$\pm$1.1} & \textbf{69.8$\pm$0.5} \\ 
        \hline
        \hline
	    
	    \end{tabular}
	    }}
	    \label{tab4}
\end{table*}

\begin{figure}[!t]
    \centering
    \setlength{\abovecaptionskip}{0.cm}
	\includegraphics[width = 1.0\textwidth]{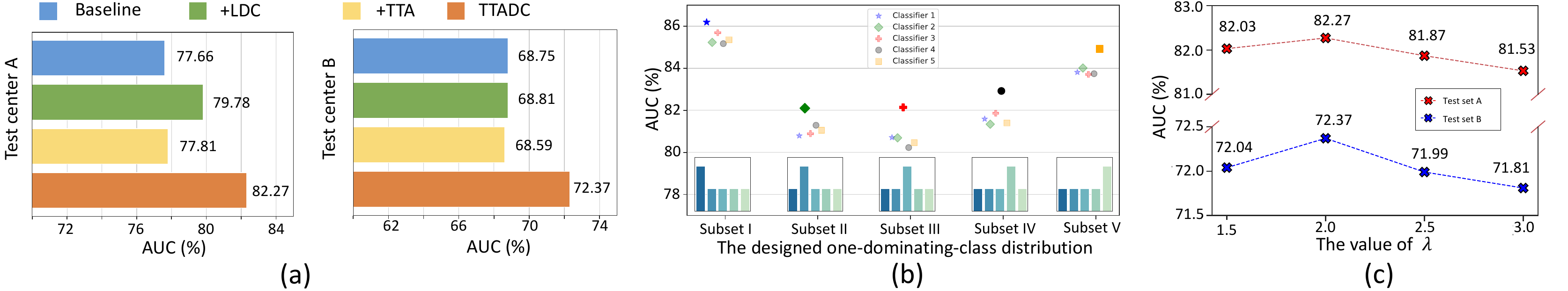}
	\caption{Ablation analysis of our method on liver CT dataset. (a) Contribution of LDC and TTA in our method; (b) Performance of our learned diverse classifiers on different one-dominating-class distributions; (c) Effect of the value of $\lambda$ on model performance. } 
	\label{abl}
\end{figure}


\subsubsection{Ablation analysis.} 
Comprehensive ablation studies have been conducted with the liver CT dataset to analyze the key ingredients regarding our TTADC.
As shown in Fig.~\ref{abl} (a), adding only the learning distribution-calibrated classifier (LDC) or test-time adaptation (TTA) over baseline is not able to improve over baseline. This is as expected since the two key components are strongly coupled, i.e., the diverse classifiers need to be properly aggregated at test time for the unknown label distribution. 
In Fig.~\ref{abl} (b), we manually sample a few images from the training center to construct the test subsets with different one-dominating-class distributions. We can see that given the test subset with $k$-th class dominating other classes, the best performance comes from the $k$-th classifier, demonstrating that our proposed distribution calibration successfully generate classifiers that have expertise on different one-dominating-class distributions. 
Moreover, Fig.~\ref{abl} (c) compares the model performance trained with different $\lambda$ in Eq.~\ref{equ4}. The results show that the optimal choice of $\lambda$ is 2.

\section{Conclusion}

We present, to our best knowledge, the first method to generalize deep classifiers to unknown test label distributions for medical image classification. Our methods innovates distribution calibration to learn multiple representative classifiers during training, which are then dynamically aggregated via test-time adaptation to deal with arbitrary label shift.
Our method is general and experiments on two important medical diagnosis tasks demonstrate the effectiveness of our method. 
\\
\\
\textbf{Acknowledgement.} This work was supported in part by the Hong Kong Innovation and Technology Fund (Project No. ITS/238/21), in part by the CUHK Shun Hing Institute of Advanced Engineering (project MMT-p5-20), in part by the Shenzhen-HK Collaborative Development Zone, in part by Jilin Provincial Key Laboratory of Medical Imaging \& Big Data (20200601003JC), Radiology, and in part by Technology Innovation Center of Jilin Province (20190902016TC).

\newpage

\footnotesize
\bibliographystyle{splncs04}
\bibliography{paper841}

%




\end{document}